\documentclass[iop,apjl,astrobib]{emulateapj}

\usepackage{natbib}

\usepackage{epsf}
\usepackage{amsbsy}
\usepackage{amssymb}
\usepackage{amsmath}
\usepackage{graphicx}

\newcommand{\be}{\begin{equation}}
\newcommand{\ee}{\end{equation}}
\newcommand{\bear}{\begin{eqnarray}}
\newcommand{\eear}{\end{eqnarray}}

\begin{document}

\lefthead{Penner et al}
\righthead{Crustal failure during binary inspiral}

\title{Crustal failure during binary inspiral} 

\author{A.J. Penner$^1$, N. Andersson$^1$,  D.I. Jones$^1$, L Samuelsson$^2$ \& I. Hawke$^1$}
\affil{$^1$School of Mathematics, University of Southampton, Southampton SO17 1BJ, UK}
\affil{$^2$Department of Physics, Ume\aa\ University, SE-901 87 Ume\aa, Sweden}

\vspace{0.2cm}
   
\begin{abstract} 
We present the first fully relativistic calculations of the crustal strain induced in a neutron star by a binary companion at the late stages of inspiral, employing realistic equations of state for the fluid core and the solid crust.  We show that while the deep crust is likely to fail only shortly before coalescence, there is a large variation in elastic strain, with the outermost layers failing relatively early on in the inspiral.  We discuss the significance of the results for both electromagnetic and gravitational-wave astronomy. 
\end{abstract}


%%%%%%%%%%%%%%%%%%%%%%%%%%%%%%%%%%%%%%%%%%%%%%%%%%%%%%%%%%%
%%%%%%%%%%%%%%%%%%%%%%%%%%%%%%%%%%%%%%%%%%%%%%%%%%%%%%%%%%%
\section{Introduction} 

Finite size effects for neutron stars in coalescing binary systems are of great interest since the departure from point-mass dynamics may provide a unique insight into the high-density equation of state when such systems are detected by future gravitational-wave experiments \citep{fh08}.
In this paper we advance our understanding of this problem by providing the first general relativistic calculations of the strain induced in the solid crust during binary inspiral, building upon our recent relativistic elastic perturbation scheme \citep{petal11}. 

The solid crust makes up a relatively small fraction of the star, and its shear modulus is small relative to the pressure.  Nevertheless, there are several reasons why an accurate treatment of the crustal strain problem is of interest.  Firstly, a sufficiently high signal-to-noise gravitational-wave observation may allow the effects of the solid phase to be identified.  Secondly, large scale failures of the crust during the  late stages of inspiral may have interesting effects, especially given the link between such mergers and short gamma-ray bursts.  Thirdly, the techniques developed here will prove essential in tackling the related problem of calculating, in General  Relativity, the maximum mass quadrupole a compact star can sustain, a problem of great interest for gravitational-wave searches (see e.g. \citet{lvc_known_pulsars_S5}).

Of particular interest is to know at what stage in the inspiral the crust fails.  Early work by \citet{koch92}  indicated this occurs only shortly before merger.  Recent estimates by \citet{ppl10} and \citet{troja} suggest failure at a much earlier stage.  We resolve this discrepancy below.  However, our modelling indicates very different levels of strain in different parts of the crust, suggesting a much richer failure sequence than simple considerations would have suggested. This is a subtlety that requires accurate calculations of the kind described here. 

%%%%%%%%%%%%%%%%%%%%%%%%%%%%%%%%%%%%%%%%%%%%%%%%%
%%%%%%%%%%%%%%%%%%%%%%%%%%%%%%%%%%%%%%%%%%%%%%%%%
\section{Back-of-the-envelope estimates}

Before discussing relativistic strained stars, we will make a few simple estimates using Newtonian theory.  As well as providing useful insights, these estimates will enable us to normalise the relativistic calculations that follow.

Consider a star of mass $M$, radius $R$, a distance $D$ from a binary companion of mass $M_{\rm comp}$.  We focus on the situation where the star is axisymmetric,  being deformed away from sphericity such that its quadrupole moment is changed by a fractional amount $\epsilon$.  This is non-zero by virtue of the tidal field of the companion, and also (we assume) by the solid phase having a zero-strain shape $\epsilon_0 \neq 0$.  This is obviously highly idealised, as in general the change in shape induced by the crustal strain will not coincide with the line joining the two stars.  However, by considering this simple case we can make some useful estimates, and make contact with other results in the literature.

The star's energy can be written  \citep{jone02}
\be
\label{eq:E_epsilon}
E(\epsilon) = E_0 + E_{\rm self-gravity} + E_{\rm tidal} + E_{\rm elastic}
\ee
where $E_0$ denotes the energy of the equivalent fluid star in the absence of a companion, $E_{\rm self-gravity}$ the perturbation in the star's self-gravity, $E_{\rm tidal}$ the energy perturbation due to the tidal interaction, and $E_{\rm elastic}$ is the elastic energy.  Since the self-gravity is minimised for a spherical configuration, we have $E_{\rm self-gravity} \sim A\epsilon^2$, where $A \sim GM^2/R$.  Assuming that the elastic energy is minimised when the star has a shape $\epsilon = \epsilon_0$, we have $E_{\rm elastic} \sim B(\epsilon-\epsilon_0)^2$, where $B$ is of order the Coulomb binding energy of the solid phase ($B \sim \int \check \mu \, dV$, where $\check \mu$ is the shear modulus).   The gravitational field of the binary partner is, of course, $-GM_{\rm comp} / r_{\rm comp}$, where $r_{\rm comp} = 0$ is the centre of mass of the companion.  Expanding this about the centre of mass of the star of mass $M$, defined by $r=0$, leads to the tidal field
\be
\label{eq:Phi_tidal}
\Phi_{\rm tidal}(r) = \frac{1}{2} r^2 \frac{GM_{\rm comp}}{D^3} [\delta_{ij} - 3n_i n_j] \hat r_i \hat r_j
\ee
where $n_i$ is a unit vector along the line separating the two stars.   The corresponding energy perturbation is
\be
E_{\rm tidal} \sim \Phi_{\rm tidal} (\epsilon M) \sim A \frac{M_{\rm comp}}{M} \left(\frac{R}{D}\right)^3 \epsilon
\ee
%(where a factor of $1/2$ has been neglected).  

Inserting  these expressions into equation (\ref{eq:E_epsilon}), minimising with respect to $\epsilon$ and making use of the expectation that $B \ll A$ leads to
\be
\epsilon = \frac{B}{A} \epsilon_0 - \frac{1}{2} \frac{M_{\rm comp}}{M} \left(\frac{R}{D}\right)^3
\ee
The first term, proportional to $\epsilon_0$, is what is normally termed the `mountain' \citep{lvc_known_pulsars_S5}, and is sourced by the star's non-spherical reference shape, caused by whatever geological processes may have been acting on it.  The second term, proportional to $M_{\rm comp}$, is sourced by the tidal field created by the companion star.  As  expected in a perturbative treatment, these two contributions to the total shape $\epsilon$ have decoupled neatly; we will denote them $\epsilon_{\rm mountain}$ and $\epsilon_{\rm tidal}$.

The corresponding perturbations in the star's quadrupole moment are found by simply multiplying by the moment of inertia $I$.  Separating the pieces we have:
\bear 
Q_{\rm mountain} &=& \frac{B}{A} \epsilon_0 I\\
Q_{\rm tidal} &=& - \frac{1}{2} \frac{M_{\rm comp}}{M} \left(\frac{R}{D}\right)^3 I 
%\sim -\frac{M_{\rm comp} R^5}{D^3} 
\label{Qtidal}
\eear

Given that it is meaningful to separate the shape changes/quadrupole moments, it must be possible to separate the strains induced in the solid phases.  In the mountain case, the actual deformation away from sphericity, $\epsilon_{\rm mountain}$, is much less than the zero strain shape $\epsilon_0$, by a factor $B/A \sim 10^{-5}$, so the strain induced is;
\be
u_{\rm mountain} \sim \epsilon_0-\epsilon_{\rm mountain} \approx \epsilon_0 \approx (A/B) \epsilon_{\rm mountain}
\label{umount}
\ee
In contrast, for the tidal deformation, the only strains induced are by the shape change $\epsilon_{\rm tidal}$ itself, so that;
\be
u_{\rm tidal} \sim \epsilon_{\rm tidal} \sim {Q_{\rm tidal}}/{I}
\label{utidal}
\ee
We see that, for a given perturbation in the quadrupole moment, a factor of $A/B \sim 10^5$ more strain is induced in the solid phases in the mountain case compared to the tidal deformation case.  Physically, one way of understanding this is as follows.  In the mountain case, the deformation is sourced by the strains themselves, and the weakness of electrostatic forces versus pressure means one needs a large strain ($\sim \epsilon_0$) to build a given mountain ($\sim \epsilon_{\rm mountain}$).  In the tidal case, the deformation is sourced by the tidal field, with the consequent shape change ($\sim \epsilon_{\rm tidal}$) being the only source of strain; the solidity of  the crust plays almost no role---the solid is deformed in much the same way as if it were fluid.

It is now straightforward to estimate at what point during inspiral the crust fails.  Combining \eqref{Qtidal} and \eqref{utidal} with
Kepler's third law, i.e.,  $\Omega^2 = GM_{\rm total}/D^3$,
%\be
%\label{eq:Omega}
%\Omega^2 = \frac{GM_{\rm total}}{D^3}
%\ee
where $\Omega = 2\pi f_{\rm orbit} = \pi f_{\rm GW}$,
%write the gravitational-wave frequency as a function of the tidal strain
%\be
%f_{\rm GW} = \left[ \frac{G}{\pi^2} \frac{2M M_{\rm total}}{M_{\rm comp}} \frac{1}{R^3} u_{\rm tidal} \right]^{1/2}
%\ee
we find that the crust fails ($u_\mathrm{tidal}=u_\mathrm{break}$) when
%\be
%f_{\rm GW, \, break} = \left[ \frac{G}{\pi^2}  \frac{2M}{R^3} u_{\rm break} \right]^{1/2}
%\ee
%so that
\be
f_{\rm GW}^\mathrm{break} \approx 3 {\, \rm kHz \,} 
\left(\frac{M}{1.4 M_\odot}\right)^{1/2} 
\left(\frac{10^6 \, \rm cm}{R}\right)^{3/2} 
\left(\frac{u_{\rm break}}{0.1}\right)^{1/2} 
\ee
(we have set $M_{\rm comp} = M$ for simplicity). We have scaled the result, which should only be trusted up to factors of order unity, to the estimated breaking strain $u_\mathrm{break}\approx 0.1$ from recent molecular dynamics simulations \citep{hk09}.
Comparing with the estimated gravitational-wave frequency at the innermost stable orbit \citep{lvccbcrates};
\be
\label{eq:f_GW}
f_{\rm GW}^\mathrm{ISCO} \approx \frac{c^3}{\pi 6^{3/2} G M_{\rm total}}  
= 1.6{\, \rm kHz \,} \left( \frac{1.4 M_\odot}{M} \right)
\ee
(again, for an equal-mass binary)
we see that crust failure should \emph{not} be expected significantly before coalescence, in agreement with the results of \citet{koch92}.

This contrasts with the estimates of \citet{ppl10} and \citet{troja} who found failure would occur much earlier in the inspiral.  By using 
the mapping between quadrupole deformation and strain appropriate for mountains (effectively equation \ref{umount}) rather than that 
for tidal deformations (as per  equation \ref{utidal}), they overestimate the strain by a factor $\sim A/B$, altering the gravitational-wave frequency at failure by a factor $\sim(A/B)^{1/2} \sim 10^2$.

%%%%%%%%%%%%%%%%%%%%%%%%%%%%%%%%%%%%%%%%%%%%%%%%
%%%%%%%%%%%%%%%%%%%%%%%%%%%%%%%%%%%%%%%%%%%%%%%%
\section{The tidal problem}

The above estimates provide a useful guide, but they do not allow us to answer important questions such as \emph{where} the crust first fails, and how large the strain \emph{variations} are within the crust.  To address these questions, we have extended the  formalism of \citet{petal11}.  As in that study, we take as our background model a spherical unstrained star (corresponding to $\epsilon_0=0$ above) and subject it to a quadrupolar field, generated by a binary companion.  However, in the present context we need to consider  the normalisation of the perturbations. Specifically, we need the mapping between the amplitude of the perturbation and the
separation, or orbital frequency, of the binary system.  The Newtonian tidal potential derived earlier can be used to fix this, as follows.

 \citet{petal11} express the metric component $g_{tt}$  in the form $g_{tt} = -e^\nu [1 - H_0(r) Y_{lm}]$, with the function $H_0$ outside 
 the star taking the form
\be
\label{eq:H_0_exterior}
H_0(r) = a_P P_{22}( r/M-1) + a_Q Q_{22}( r/M-1)
\ee
where $r$ is the radial distance from the centre of the star, $P_{22}$ and $Q_{22}$  are the associated  Legendre functions, and $a_P$, $a_Q$ are constants.  
%Note that far from the star, i.e. $x  \gg 1$, Far from the star, if we retain the leading order term from each part we find: 
%\be 
%H_0(r \gg R) \approx 3 a_P \left(\frac{r}{M}\right)^2 +  \frac{8}{5}  \left(\frac{r}{M}\right)^{-3} a_Q
%\ee
%The first term can be identified with the tidal field provided by the companion, the second term the change in field due to the quadrupolar response of the star.
%\be
%P_{lm} \sim x^l, \hspace{10mm} Q_{lm} \sim x^{-(l+1)}
%\ee
Far from the star the dominant term is
\be
H_0(r \gg R) \approx  3 a_P \left(\frac{r}{M}\right)^2 
\ee
In this limit the metric can be related to the Newtonian potential according to
\be
\Phi^{\rm Newton} = -\frac{g_{tt}+1}{2}
\ee
This allows us to relate the asymptotic form of $H_0$ to the Newtonian tidal potential in equation (\ref{eq:Phi_tidal}), leading to
\be
\label{eq:a_P}
a_P = \frac{2}{3} \sqrt{\frac{4\pi}{5}} \frac{M_{\rm comp}M^2}{D^3} .
\ee
The numerical calculation can then be normalised by considering the value of $H_0$ at the stellar surface:
\be
\label{eq:H_0_surface}
H_0(R) = a_P [P_{22}( R/M-1) + a_2 Q_{22}( R/M-1)]
\ee
where the ratio $a_2 = a_Q/a_P$ is related to the Love number $k_2$ \citep{fh08,hllr10} by:
\be
\label{eq:a_2}
a_2 = \frac{15}{4} \left(\frac{R}{M}\right)^5 k_2
\ee
The Love number is calculated numerically as a property purely of the stellar model, and the amplitude of the associated perturbation is parameterised by $a_P$, either in terms of the binary separation, or the associated gravitational-wave frequency:
\be
\label{eq:a_P_dim_f}
a_P = {6\pi^2} \sqrt{\frac{4\pi}{5}} \frac{G^2 M^2 M_{\rm comp}}{c^6 M_{\rm total}} f_{\rm GW}^2
\ee
Note that, for given stellar masses the tidal deformation scales as the square of the gravitational-wave frequency.

%%%%%%%%%%%%%%%%%%%%%%%%%%%%%%%%%%%%%%%%%%%%%%%%%
%%%%%%%%%%%%%%%%%%%%%%%%%%%%%%%%%%%%%%%%%%%%%%%%%
\section{Results}

We have generalised the numerical framework of \citet{petal11} to allow for realistic equations of state for both the crust and the star's core.  We present results for stellar models that combine the \citet{akmal} equation of state for the core fluid with the results of \citet{dh01} in the crust. These models are state-of-the-art for this problem, but it should be noted that we have not accounted for the (likely) presence of nuclear pasta in the inner crust. A sizeable pasta region could have significant impact on the results, but we do not yet have a sufficiently detailed equation of state representing this possibility. 

\begin{figure}[h]
\setlength{\unitlength}{1mm}
\hskip  0mm
\begin{center}
\includegraphics[width=80mm,angle=0,clip]{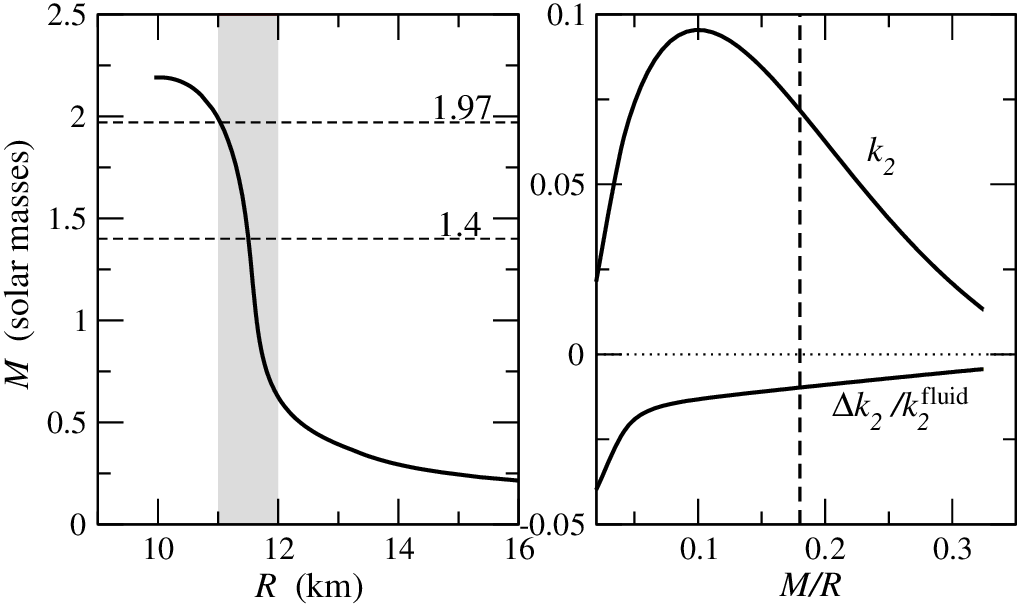}
\end{center}
\caption{Left panel: Mass-radius relation for our sequence of stellar models, demonstrating consistency with the observational constraints of \citet{detal10} (upper dashed horizontal line) and \citet{slb10} (grey region).   
Right panel: The Love number $k_2$ (upper curve)  as a function of the stellar compactness $M/R$; cf., the pure fluid results in Fig.~1 of \citet{hllr10}.  We also show the relative influence of the crust on the tidal deformability, represented by $\Delta k_2/k_2^\mathrm{fluid}$ (lower curve); this is similar to the  results of \citet{petal11}. The compactness of the $1.4M_\odot$ model considered in Fig.~\ref{fig:strain} is indicated by a vertical dashed line.}
\label{fig:mass_and_love}
\end{figure}

The chosen core equation of state is sufficiently stiff to satisfy constraints from observations (cf., the left panel of Fig~\ref{fig:mass_and_love}). It allows for neutron star masses  
at least as large as $2 M_\odot$, in agreement with the observed $1.97M_\odot$ mass  of PSR J1614-2230 \citep{detal10}.  
It also satisfies the radius constraint from  X-ray burst sources, i.e., that a star with mass of $1.4M_\odot$ should have a radius in the range 11--12~km \citep{slb10}. The elastic properties of the crust do not affect the equilibrium configurations since we assume the  star is relaxed at large binary separations.

Given this equilibrium configuration, we have calculated both the Love number $k_2$, and the fractional difference  $\Delta k_2/k_{2}^\mathrm{fluid}=(k_2^\mathrm{crust}-k_{2}^\mathrm{fluid})/k_{2}^\mathrm{fluid}$ between the Love numbers for an elastic star and the equivalent purely fluid star (see \citet{petal11} for details). The results are shown in the right panel of Fig.~\ref{fig:mass_and_love}.  They can be compared to, first of all, the fluid star results of \citet{hllr10} and secondly the results of \citet{petal11} for the magnitude of the crust effects. Based on these comparisons, the present results are not  surprising. 

\begin{figure}[t]
\setlength{\unitlength}{1mm}
%\hskip  0mm
\begin{center}
\includegraphics[width=80mm,clip]{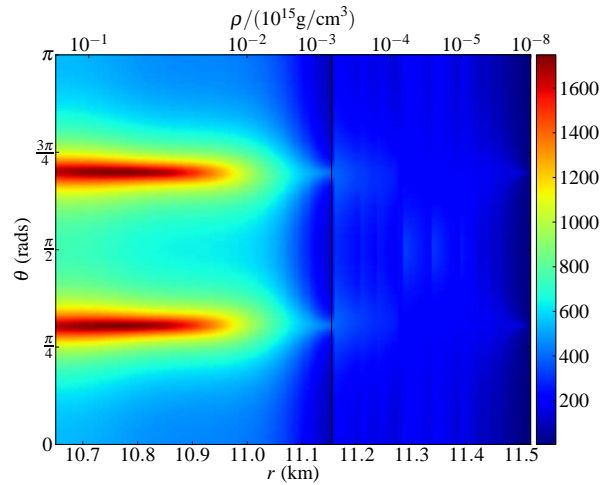}
\end{center}

\caption{The gravitational-wave frequency (in Hz) at failure for different locations in the crust of a $1.4M_\odot$ neutron star. The result, which is obtained by comparing the von Mises strain to the breaking strain of \citet{hk09}, corresponds to an equal-mass binary.
The vertical  line near 11.15~km indicates the  location of
neutron drip in the star.}
\label{fig:strain}
\end{figure}

We have also evaluated the von Mises stress associated with the tidal perturbation. By comparing the result to the anticipated breaking strain
$u_\mathrm{break}\approx 0.1$ \citep{hk09}, 
we can infer when different parts of the crust fail during binary inspiral. A typical result is shown in Fig.~\ref{fig:strain}, providing the gravitational-wave frequency at failure throughout the crust for a $1.4M_\odot$ star (in an equal-mass binary). The result is not trivial, owing to both the nonlinear  combination of perturbed quantities that enter the von Mises stress and the associated angular functions.  

The large variation in the crustal strain implies that failure will occur at different stages during inspiral. From  Fig.~\ref{fig:strain}
we see that  the outer crust (roughly up to neutron drip, corresponding to $r \approx 11.15$~km in Fig.~\ref{fig:strain}) fails fairly uniformly when  $f_{\rm GW}\approx 200$~Hz. Meanwhile, failure of the bulk of the inner crust requires $f_{\rm GW}\approx 600-800$~Hz, a factor of two or so below the ISCO frequency (see equation (\ref{eq:f_GW})).  However, there are also macroscopic regions in the inner crust that will not fail before merger. These results improve upon the estimates from section~2, showing the rich structure of the realistic calculation, with failure occurring at different stages at different depths.

%%%%%%%%%%%%%%%%%%%%%%%%%%%%%%%%%%%%%%%%%%%%%%%%%%%%%
%%%%%%%%%%%%%%%%%%%%%%%%%%%%%%%%%%%%%%%%%%%%%%%%%%%%%
\section{Implications}

What happens when the crust fails? Again, key insights are provided by the molecular dynamics simulations of \citet{hk09}. 
The indications are that when the critical strain is reached, there is a catastrophic failure, with energy released throughout the strained volume, rather than the formation of a  lower-dimensionality crack.  What happens next is less clear.  Two extreme scenarios can be envisaged: the relieved strain is dissipated locally as heat, or, the strain is converted into phonons/seismic waves and transported throughout the star prior to dissipation.  An interesting question is whether or not the relieved strain is capable of melting the crust.  From the estimates in Section~2 we see that the strain energy may be at the level of $10^{46}$~erg. However, it is easy argue that  even the release of this level of energy will not have significant impact on the crust.   To demonstrate this, let us  assume that the first of the two extreme cases described above applies, with all strain energy being dissipated locally.

Ignoring the temperature dependence of the shear modulus (which would be irrelevant for typical neutron star temperatures anyway),  the crust melts when $\Gamma \lesssim 173$, where \citep{melting}
\be
\Gamma = {Z^2 e^2 \over a} {1 \over k_B T} 
\label{Gamma}
\ee
Here, $e$ is the electron charge, $Z$ is the charge per ion and $a=\left(3/4\pi n_i \right)^{1/3}$, with $n_i$  the ion number density, is the average ion spacing.
For a typical temperature of $10^8$~K, we have $\Gamma \sim 10^4$ in the inner crust (where $Z\approx 50$ and the number of nucleons per ion, $A$, is several hundred).

We want to compare the melting temperature obtained from Eq. ~\eqref{Gamma} to the temperature reached after releasing the tidal strain.  For this we need the shear modulus, which is  approximated by  \citep{shearmod}
\be
\check{\mu} \approx 0.1 n_i {Z^2 e^2 \over a}
\ee
The energy per unit volume corresponding to a strain $u$ is then
\be
E_\mathrm{strain} \approx10^{-2}  \check{\mu} \left( \frac{u}{0.1} \right)^2
\ee
We can turn this  into a temperature via the  heat capacity, $T_\mathrm{strain} \approx {E_\mathrm{strain} /c_v}$,  
 making use of the estimate of the  heat capacity of ions from \citet{vanriper};
$C_v = {c_v /n_i} = \alpha k_B$
where $\alpha$ is of order unity (for $10^8$~K we have $\alpha \approx 5$). Thus, we see that  the 
released heat will melt the crust if
\be 
%  \left( {u_\mathrm{break} \over 0.1 } \right)^2 > 5 \alpha
u  \gtrsim 0.5 \left(\frac{\alpha}{5}\right)^{1/2}
\ee
This exceeds the breaking strain estimates of \citet{hk09}, indicating that  the released heat does \emph{not} melt the crust during binary inspiral, even when \emph{all}  the strain energy goes into local heating.

\begin{figure}[h]
\setlength{\unitlength}{1mm}
%\hskip  15mm
\begin{center}
\includegraphics[width=65mm,clip]{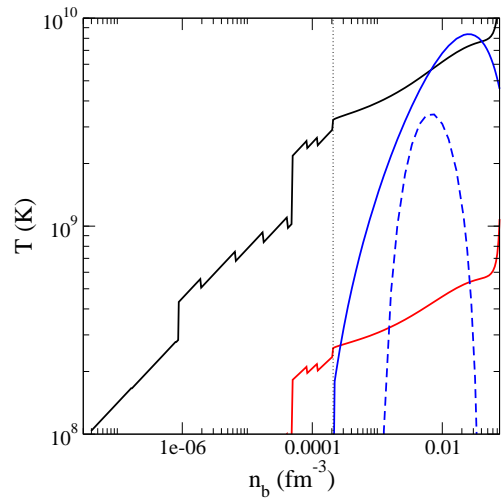}
\end{center}

\caption{Various relevant temperatures  as functions of baryon number density $n_b$ throughout the crust.  The upper curve is the crustal melting temperature, the lower curve the temperature to which the crust would be heated by local dissipation of strain energy into heat (cf., the discussion in the text), demonstrating that the crust does not melt during inspiral.  We also show two models for the neutron (singlet) critical temperature (corresponding to models a (solid) and c (dashed) of \citet{nuclpa}), showing that the local heating is insufficient to break superfluidity. The vertical  line indicates the  location of
neutron drip. }
\label{fig3}
\end{figure}

A more detailed calculation leads to the results in Figure~\ref{fig3}, which compares the melting temperature to that reached after releasing the strain locally. We also compare to the critical temperature for (singlet) neutron superfluidity. The message from this comparison is clear; the  heat that is released when the crust fails is \underline{not} sufficient to break the superfluid pairing, either.  This implies that a realistic numerical treatment of neutron star binary inspiral should allow for superfluidity and crust elasticity prior to coalescence.
 
The crust may not melt, but we are still releasing a significant amount of energy when the crust fails. Could this have observable consequences? Given the anticipated importance of neutron star binary mergers for future  gravitational-wave observations, and the likely association of such events with short gamma-ray bursts, a precursor signal would undoubtedly be interesting. Observational evidence for such precursors was recently discussed by \citet{troja}.
However, the heating of the crust should be irrelevant in this respect.  The results in Fig.~\ref{fig3} show that the outer crust does not heat significantly. The inner crust  heats up, but the associated heat would not diffuse to the surface before the binary merges (see \citet{blaes} for discussion). The possibility that the energy is released into seismic waves that generate Alfv\'en waves in the magnetosphere, eventually leading to a gamma-ray signal, is more interesting (again, see \citet{blaes}). In principle, such an event could be as energetic as the largest observed magnetar flare (the 27 December 2004 event in SGR 1806-20 \citep{palmer}), assuming that the entire strain energy is transferred to the magnetosphere. Such a signal could possibly be observable from a distance of 100~Mpc. The corresponding gravitational-wave signal would be comfortably detectable by the advanced LIGO/Virgo detector network. However, according to the rate estimates of \citet{lvccbcrates} there is  likely to be only of order one such event per year of observation. According to our numerical results, the most energetic failure event would precede the merger itself by a fraction of a second. Observations of such precursors would obviously be tremendously interesting, but it seems clear from our results that they would have to be associated with rather unique events.

Finally, it remains to consider the impact on the gravitational-wave signal.  The smallness  of the effect of the elastic crust on the tidal distortion (right panel of Fig.~\ref{fig:mass_and_love}), suggests it would require an unfeasibly high signal-to-noise observation for the crustal elasticity to be apparent, especially given that finite-size effects themselves are likely to be only borderline detectable by the next generation interferometers \citep{hllr10}.  However, as we have shown, even at large separations, portions of the crust will fail, with the volume of crust pushed into failure steadily increasing as the inspiral proceeds.  \citet{koch92} suggested that this may lead to anomalous frictional damping, which may have a relevant secular affect on the phasing.    A proper  calculation of this would take the strain field presented here as its starting point,  but would  incorporate a more realistic treatment of crust failure, going beyond the simple breaking strain criterion to allow for a regime of plastic deformation and dissipation.     An accurate study of this effect is clearly needed to assess its significance for gravitational-wave searches.  This would be an important but non-trivial  extension of the work presented here.

\acknowledgements

AJP acknowledges support from a VESF fellowship. 
NA, IH and DIJ  acknowledge support from STFC via grant no. PP/E001025/1.  
LS is supported by the European Research Council under contract no. 204059-QPQV, and the Swedish Research
Council under contract no. 2007-4422.  
We acknowledge support from CompStar (an ESF Research Networking Programme).

%%%%%%%%%%%%%%%%%%%%%%%%%%%%%%%%%%%%%%%%

\end{document}